\documentclass[10pt,letterpaper]{article}
\usepackage{amsmath}
\usepackage{amssymb}
\usepackage{cite}
\usepackage{color}
\usepackage{opex3}
\usepackage{txfonts}
\usepackage{epstopdf}
\usepackage{amssymb}
\usepackage{tipa}
\usepackage{graphicx}
\usepackage{txfonts}
\usepackage{amssymb}
\usepackage{extarrows}

\begin{document}

\title{Single-photon transport in one-dimensional coupled-resonator waveguide with second order nonlinearity coupling to a nanocavity containing a two-level atom and Kerr nonlinearity}
\author{Hongyu. Lin$^{1,2}$, Xiaoqian. Wang$^{1}$, Zhihai. Yao$^{1,4}$ and Dandan Zou$^{3,5}$}
\address{
$^1$ Department of Physics, Changchun University of Science and Technology, Changchun - 130022, China \\
$^2$ College of Physics and electronic information, Baicheng Normal University, Baicheng - 137000, China\\
$^{3}$ School of Physics and Electronic Information, Shangrao Normal University, Shangrao - 334001, China\\
$^4$yaozh@cust.edu.cn\\
$^5$37949126@qq.com\\
}

\begin{abstract}
We study controllable single photon scattering in a one-dimensional waveguide coupling with an additional cavity by second order nonlinear materials in a non-cascading configuration, where the additional cavity is embedded with two-level atom and filled with Kerr-nonlinear materials. Considering the second order nonlinear coupling, we analyze the transmission properties of the three different coupling forms as follows: (i) The two-level atom is excited without the Kerr-nonlinearity. (ii)The Kerr-nonlinearity is excited without the two-level atom. (iii) Both of the two-level atom and Kerr-nonlinearity are excited. The transmission and reflection amplitudes are obtained by the discrete coordinates approach for the three cases. The results showed that the transmission properties can be adjusted by the above three different coupling forms, which indicate our scheme can be used as a single photon switch to control the transmission and reflection of the single photon in the one-dimensional coupled resonant waveguide.
We compared the results with [Phys. Rev. A 85, 053840(2012)] and find the advantages.
\end{abstract}






\section{Introduction}
Single photon are considered as one of the most suitable carriers for quantum information. The single-photon control, which plays a key role in the realization of quantum information processing and quantum communication~\cite{1}. In the past decades, many systems choose a atom as stationary qubits to achieve strong photon-atom interaction in a high-quality optical microcavity~\cite{2,3,4}.
Recently, single-photon transmission has been widely used in quantum networks~\cite{5}, optical circuits~\cite{6,7}, quantum switches~\cite{8,9,10,11,12,13} and high-precision spectroscopy~\cite{14}, which have been extensively investigated both theoretically~\cite{15,16,17,18,19,20} and experimentally~\cite{21,22,23}.
The interaction between nanosphotonic waveguide and matter or additional cavities makes a single photon control approach with wide application prospects, and the waveguides~\cite{24,25,26} and nanofibers~\cite{27,28,29,30,31,32,33} can well simulate one-dimensional light propagation~\cite{34,35,36,37,38,39,40}. The coupled resonator waveguide (CRW) is an important and widely used one-dimensional optical waveguide model. This can be achieved through photonic crystals~\cite{45} or a wire resonator using a superconducting material for transmissions~\cite{46,47,48}. In the CRW, We can control the transmission and reflection of a single photon by the coupling of CRW with the additional cavity, and obtaining strong cavity interaction, thus affecting the scattering properties of a single-photons. In addition, single-photon transmission realized by local and non-local coupling with defects in one-dimensional discrete systems has also been studied, which mainly discusses nonlinear Fano resonance and nonlinear Fano-Feshbach resonance~\cite{49,50,51}. At present, photon scattering in multi-level quantum emitter systems is also mentioned~\cite{l1,l2}.

The second-order nonlinear system is an important and widely used optical system~\cite{52}, which can convert a single photon in a high frequency cavity into a two-photon in a low frequency cavity by parametric down conversion, and the reverse process can also be realized by parametric up conversion~\cite{53,54}. Many materials can achieve second-order nonlinearity, such as III-V semiconductors (e.g., GaAs, GaP, GaN, AlN, etc.)~\cite{55,56}, and in recent years, the use of the LiNbO3 waveguide as a high high-$\chi^{(2)}$ nonlinear material platform has also been widely studied~\cite{ll1,ll2,ll3}, which have a wide range of applications, such as strongly coupled single photon~\cite{57}, anticlustering of photons~\cite{58,59}, and optically induced transparency~\cite{60}.

In this paper, we introduce the second-order nonlinearity, into the CRW to realize the quantum optical switch.
We study controllable single photon scattering in a one-dimensional waveguide coupling with an additional cavity by second order nonlinear materials, where the additional cavity is embedded with two-level atom and filled with Kerr-nonlinear materials. Three cases are considered: (i) The two-level atom is excited without the Kerr-nonlinearity. (ii)The Kerr-nonlinearity is excited without the two-level atom. (iii) Both of the two-level atom and Kerr-nonlinearity are excited. By the discrete coordinates approach we obtain the reflection and transmission rate in the current system. The effects of second-order nonlinearity, Kerr nonlinearity and two-level atoms acting as a single-photon switch to control the transmission and reflection in a one-dimensional coupled resonant waveguide are analyzed, respectively. The results show that in these three systems both of the perfect transmission and reflection areas can be controlled by adjusting the system parameters. The contribution of the present scheme can be summarized as: (i)For the first time introduce the Kerr-nonlinearity into the CRW to realize the quantum optical switch. (ii) In comparison with the CRW coupling to a two-level system~\cite{aa}, which find that under the same parameters, the second-order nonlinear system can forms two perfect reflection regions, it is different from the perfect reflection dips formed by the two-level system. (iii) The case of both of the two-level atom and Kerr-nonlinearity are excited in the system, the Kerr nonlinearity can also be used as a single photon switch to control a single photon to realize transmission and reflection.

The manuscript is organized as follows:
In Sec.~{\rm II}, we introduce the physical model and the theory for the system, where the analytic solutions of transmission and reflection amplitudes are obtained.
In Sec.~{\rm III}, the controllable scattering properties are studied in the case of only the two-level atom is excited, and comparison with the CRW coupling to a two-level system.
In Sec.~{\rm IV}, we study the effects of Kerr nonlinearity on the scattering properties of single photons in this system.
Discussion and conclusion are given in Sec.~{\rm V}.

\section{Physical model}

\begin{figure}[h]
\centering
\includegraphics[scale=0.80]{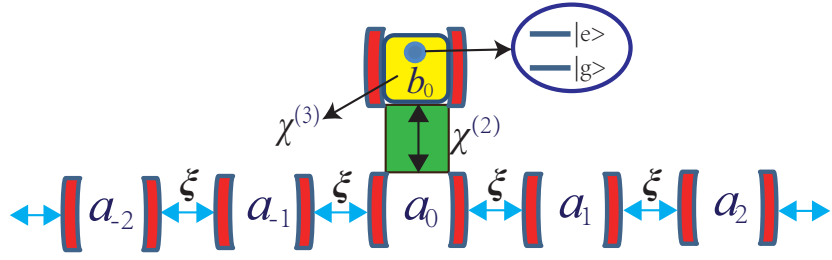}
\caption{(Color online)
Schematic illustration of the coherent transport of a single photon in a coupled-resonator waveguide coupling with an additional cavity by second order nonlinear materials in a non-cascading configuration, where the additional cavity is embedded with two-level atom and filled with Kerr-nonlinear materials.}
\label{zhuangzhi}
\end{figure}
We consider a hybrid system, where the system contains a CRW coupling with an additional cavity by second order nonlinear materials in a non-cascading configuration, where the additional cavity is embedded with two-level atom and filled with Kerr-nonlinear materials. which is illustrated in Fig.~\ref{zhuangzhi}. The Hamiltonian of the system can be written as
\begin{eqnarray}
H&=&H_a+H_b+H_g+H_e+H_J+H_U,
\label{01}
\end{eqnarray}
here $H_a$ expresses the a single photon transport in the
CRW, $H_b$ and $H_e$ describes the additional resonator and the two-level atom, respectively, $H_g$ denotes
the interaction between the CRW and the additional resonator, $H_J$ denotes interaction between the additional resonator and two-level atom, and the $H_U$ expresses the Kerr-nonlinearity. The six terms are given as (we set $\hbar=1$ hereafter)
\begin{eqnarray}
&&H_a=\omega_a\sum_ja_j^{\dag}a_j-\xi\sum_j(a_j^{\dag}a_{j+1}+a_{j+1}^{\dag}a_{j}),\nonumber\\
&&H_b=\omega_bb_0^{\dag}b_0,\nonumber\\
&&H_g=g(a_0^{\dag}b_0^2+b_0^{\dag2}a_0),\nonumber\\
&&H_e=\omega_e\sigma^{\dag}\sigma,\nonumber\\
&&H_J=J(b_0^{\dag}\sigma+\sigma^{\dag}b_0),\nonumber\\
&&H_U=Ub_0^{\dag}b_0^{\dag}b_0b_0,
\label{02}
\end{eqnarray}
where $a_j$ ($a_j^{\dag}$) is the annihilation (creation) operator of the $j$th cavity in the CRW, $\omega_a$ denotes the resonance frequency of the cavities, and the coefficient of $\xi$ is the hopping energy between the two nearest-neighbor resonators
of the CRW. $b_0$ ($b_0^{\dag}$) is the annihilation (creation) of the additional cavity with frequency $\omega_b$. The $g$ is the coupling strength between the $0_{th}$ resonator and the additional cavity, which mediates the conversion of the photon in cavity $a_0$ into two photons in cavity $b_0$. $\sigma$ ($\sigma^{\dag}$) is the annihilation (creation) of the atom in the additional resonator $b_0$ with frequency $\omega_e$, and $J$ is the coupling strength between the atom and the additional cavity. $U$ is the three-order nonlinear strength.

In this system the number of excitations is conserved, and restrict it to the one-excitation subspace for the CRW, the single-photon transports from the left of the CRW to the resonator $a_0$, the photon will come to the resonator $b_0$ by second-order nonlinear system and be converted into two photons. Next, the two photons also interact with the atom in the resonator cavity $b_0$, and the Kerr nonlinear materials will also have a certain effect of the two photons. While the two photons in resonator $b_0$ come back to resonator $a_0$, they will be converted to one by the second-order nonlinearity. So, this guarantees that there is only one photon in the CRW. For single excitation subspace for the CRW, the eigenstate of the system has the form
\begin{eqnarray}
|E_k\rangle=\sum_ju_k(j)a_j^\dag|0\rangle_a|0\rangle_b|g\rangle + u_b|0\rangle_a|2\rangle_b|g\rangle + u_e|0\rangle_a|1\rangle_b|e\rangle,
\label{03}
\end{eqnarray}

Where $|0\rangle_a|0\rangle_b|g\rangle$ indicates the vacuum states of the CRW or resonator cavity $b_0$, and the two-level atom is in the ground state. The $|0\rangle_a|2\rangle_b|g\rangle$ indicates two-photon state of the resonator cavity $b_0$ and the vacuum state of the CRW, and the two-level atom is in the ground state. The $|0\rangle_a|1\rangle_b|e\rangle$ indicates the two-level atom is in the excited state and one-photon state of the resonator cavity $b_0$, and the vacuum state of the CRW. $u_b$ and $u_e$ are the corresponding amplitudes, respectively.
According to the steady state schr\"{o}dinger equation $H|E_k\rangle=E_k|E_k\rangle$, we get a set of equations for the coefficients
\begin{eqnarray}
(E_k-\omega_a)u_k(j)-\sqrt{2}gu_b\delta_{j,0}&=&-\xi[u_k(j-1)+u_k(j+1)],\nonumber\\
(\omega_e+\omega_b-E_k)u_e&=&-\sqrt{2}Ju_b\nonumber\\
(2\omega_b+2U-E_k)u_b&=&-\sqrt{2}Ju_e-\sqrt{2}gu_k(0).
\label{04}
\end{eqnarray}
By eliminating the amplitudes $u_b$ and $u_e$, we can obtain the discrete-scattering equation
\begin{eqnarray}
(E_k-\omega_a+V_{eg})u_k(j)=-\xi[u_k(j-1)+u_k(j+1)],
\label{05}
\end{eqnarray}
here
\begin{eqnarray}
V_{eg}=-\frac{2g^2\delta_{j,0}(E_k-\omega_e-\omega_b)}{(E_k-2\omega_b-2U)(E_k-\omega_e-\omega_b)-2J^2}.
\label{06}
\end{eqnarray}
The effective potential $V_eg$, resulting from the interaction between the additional
resonator and the CRW located at site $j = 0$, modifies the single-photon
transport property in the resonator waveguide. When the coefficient $g$ of second-order nonlinearity equals to zero, the effective potential $V_eg$ vanishes.
We assume that the single photon is injected from the left side of the CRW, then a usual solution for the scattering equation is
\begin{equation}
u_k(j)=
\left\{
 \begin{array}{lr}
   e^{ikj}+re^{-ikj}, & j<0, \\
     te^{ikj}, & j>0,
  \end{array}
 \right.
 \label{07}
 \end{equation}
here the $t$ is the transmission amplitudes, and the $r$ is the reflection amplitudes. And the $e^{ikj}$ denotes the wave traveling to the right, and $e^{-ikj}$ denotes the wave traveling to the left side.
Only consider the CRW, $E_k$ is characterized by
\begin{eqnarray}
E_k=\omega_a-2\xi \cos k.
\label{08}
\end{eqnarray}
By considering the continuity condition $u_k(0^+)=u_k(0^-)$ and Eqs.~(\ref{05})-(\ref{08}), the transmission and reflection amplitude equations can be writed as
\begin{eqnarray}
t&=&\frac{2i\xi [(E_k - 2 \omega_b-2U)(E_k-\omega_e-\omega_b)-2J^2] \sin k}{-2 g^2(E_k-\omega_e-\omega_b) + 2 i \xi[(E_k - 2 \omega_b-2U)(E_k-\omega_e-\omega_b)-2J^2] \sin k},\nonumber\\
r&=&\frac{-2 g^2(E_k-\omega_e-\omega_b)}{-2 g^2(E_k-\omega_e-\omega_b) + 2 i \xi[(E_k - 2 \omega_b-2U)(E_k-\omega_e-\omega_b)-2J^2] \sin k}.
\label{09}
\end{eqnarray}
where $\Delta_a \equiv E_k - \omega_a$, $\Delta_b \equiv E_k - 2 \omega_b$ and $\Delta_{eb} \equiv E_k-\omega_e-\omega_b$, we rewrite the $t$ of Eq.~(\ref{09}) in another form
\begin{eqnarray}
t&=&\frac{2i\xi [(\Delta_b-2U)\Delta_{eb}-2J^2] \sqrt{1-(\frac{\Delta_a}{2\xi})^2}}{-2 g^2\Delta_{eb} + 2 i \xi[(\Delta_b-2U)\Delta_{eb}-2J^2] \sqrt{1-(\frac{\Delta_a}{2\xi})^2}},
\label{10}
\end{eqnarray}
Here the transmission rate is $T=|t|^2$, the reflection rate is $R=|r|^2$. The relation $T+R=1$ can be easily verified, so we only analyze the transmittance $T$ of a single photon in the system. In this paper, we use a cyclic three-level artificial atom of a superconducting flux quantum circuit interacting with a two-mode superconducting transmission-line resonator to realize the second-order nonlinear system~\cite{ff1}, where the second-order nonlinear coupling strength $g$ can be adjusted by adjusting the position of the qutrit in the transmission-line resonator, and the value of $g/(2\pi)$ can reach 5MHz. The CRW can also be realized by using
coupled superconducting transmission line resonators~\cite{8}.
Experimentally, large-scale ultrahigh-Q
coupled cavity arrays based on the transmission line resonators have
been realized~\cite{shiyan0}.

\section{Scattering properties of a single photon inside a CRW with second-order nonlinearity and two-level atom}

Next, we will study the single-photon scattering properties in the one-dimensional CRW with second-order nonlinearity and two-level atom, and we set the parameters $U=0$ in this section. For convenience, all parameters are in units of $\xi$ in this paper.
\begin{figure}[h]
\centering
\includegraphics[scale=0.80]{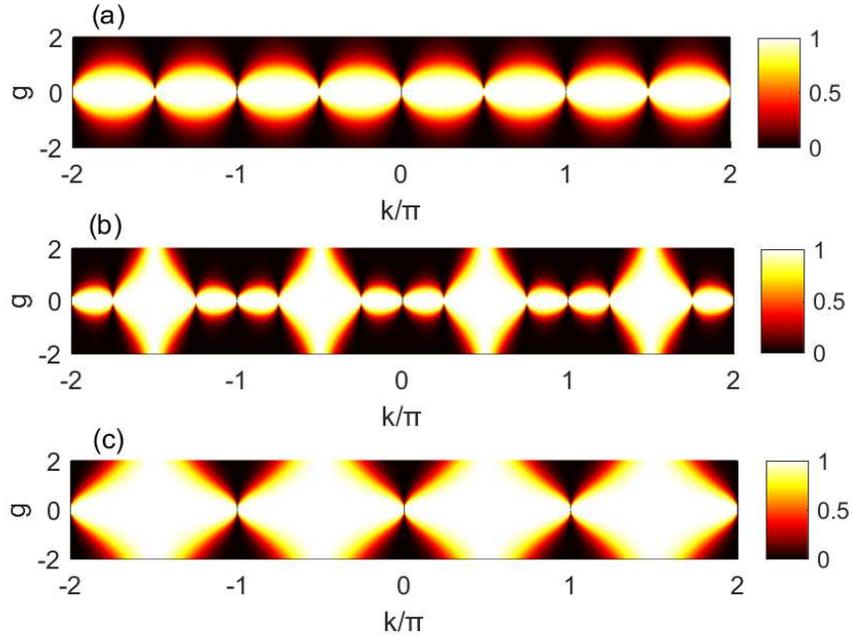}
\caption{(Color online) Transmission rate $T$ as a function of $k$ and $g$ with different $J$. $\omega_a=6$, $\omega_b=\omega_e=3$ and $U=0$.(a) $J=0.1$, (b) $J=1$, (c) $J=2$. All the  parameters are rescaled by the hopping energy $\xi$.
 }
\label{fig2}
\end{figure}

In Fig.~\ref{fig2}, we plot the transmission rate $T$ as a function of $k$ and $g$ with different $J$, and the parameters $\omega_a=2\omega_b=2\omega_e=6$ are same in all the three pictures.
In Fig.~\ref{fig2}(a), we set the $J=0.1$, the shape of $T=1$ looks like a oval areas of the fore-and-aft link. The joint positions of the oval areas can be calculated as $k=n\pi$ or $k=\arccos\frac{\sqrt{2}J}{2\xi}$. If one of the joint position conditions is satisfied, the nodes appear. In Fig.~\ref{fig2}(b), we set parameters $J=1$, as the value of $J$ increases, large rhomboid transmission areas occurs periodically at $k=\arccos(\frac{\sqrt{2}J}{2\xi})$, and in addition to the rhomboid transmission regions, the transmission regions controlled by parameter $g$ becomes smaller. In Fig.~\ref{fig2}(c), we set $J=2$, as the value of $J$ goes up, when $\frac{\sqrt{2}J}{2\xi}>1$ the position conditions $k=\arccos\frac{\sqrt{2}J}{2\xi}$ doesn't work, the nodes only depends on $k=n\pi$, at the same time, the whole system shows a strong transmission property. So, when the parameter $g$ is determined, the $J$ can act as a single photon switch to control the single-photon transmission and reflection in the one-dimensional coupled resonator waveguide.

\begin{figure}[h]
\centering
\includegraphics[scale=0.80]{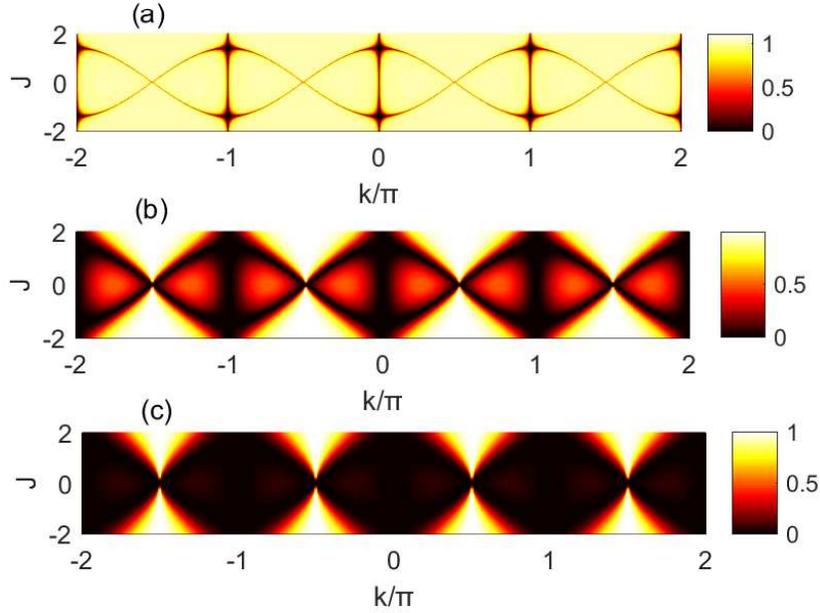}
\caption{(Color online) Transmission rate $T$ as a function of $k$ and $J$ with different $g$. $\omega_a=6$ and $\omega_b=\omega_e=3$.(a) $g=0.1$, (b) $g=1$, (c) $g=2$. All parameters are in units of $\xi$ in this paper.}
\label{fig3}
\end{figure}

In Fig.~\ref{fig3}, we plot the transmission rate $T$ as a function of $k$ and $J$ with different $g$, same as the Fig.~\ref{fig2} the parameters $\omega_a=2\omega_b=2\omega_e=6$ are same in all the three pictures. In Fig.~\ref{fig3}(a), we set the photon to come in the CRW from the left, and the parameters $g=0.1$, since the second order nonlinear strength is weak, so, the single photon can pass through the cavity $a_0$ without the impact of destructive interference. we can get the same conclusion from the expression for the effective potential $V_eg$ show in Eqs.~(\ref{06}).
As the value of parameter $g$ increases, the transmission areas is significantly reduced, the switching action of parameter $g$ is reflected. The same conclusion with Fig.~\ref{fig3}, when the parameter $J$ is determined, the $g$ can also act as a single photon switch to control the single-photon transmission and reflection in the one-dimensional coupled resonator waveguide.

\begin{figure}[h]
\centering
\includegraphics[scale=0.50]{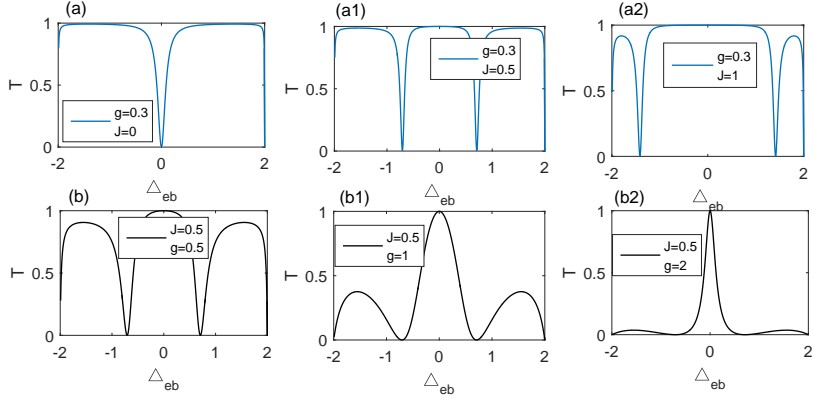}
\caption{Transmission rate $T$ as a function of $\Delta_{eb}$ with different $g$ and $J$. In the calculation, $\omega_a =2\omega_b= 2\omega_e=4$, and all the parameters are rescaled by the hopping energy $\xi$.
  }
\label{fig4}
\end{figure}

In order to further investigate the scattering properties of the second-order nonlinearity coupled system, and the coupling strengths  $g$ and $J$ are important parameters on the influence of photon transmission properties. We show $T$ as a function of $\Delta_{eb}$ with different $g$ and $J$ in Fig.~\ref{fig4}, and we set the third-order nonlinear strength $U=0$ at present.

In fig.~\ref{fig4}(a), (a1) and (a2) the second order nonlinear strength $g=0.3$ are same, in Fig.~\ref{fig4}(a) we set $J=0$, which means that the two level atom is not involved in system coupling. A dip happens at $\Delta_{eb}=0$ in the transmission spectrum, and the case of $J=0$ incident that $\omega_e=0$, the point where $\Delta_{eb}=0$ means $E_k=\omega_b$, and this indicates the cavity $b_0$ and the photon are resonant states in this case, so, the single-photon is perfect reflected at the site $\Delta_{eb}=0$. In Fig.~\ref{fig4}(a1), we set $J=0.5$, and the two level atom be excited, there are two points in the transmission spectrum where $T=0$, according to the transmission amplitude equation Eqs.~(\ref{10}) and the given parameters, the width between the two dips $W=2\sqrt{2}J$, so as the value of $J$ increases, the width of the two optimal reflection points increases the results are shown in fig.~\ref{fig4}(a2). If the coupling strength $J$ further increases such that $J \gg g$ , the system becomes nearly transparent. In fig.~\ref{fig4}(b), (b1) and (b2) we plot the transmission rate $T$ as a function of $\Delta_{eb}$ with different $g$, and the parameter $J=0.5$ are same in these three pictures. It is evident from the three transmission spectrums, the peak at $T=1$ becomes narrow and the two side peaks are strongly suppressed as the increase of $g$. As the value of $g$ further increase the two side peaks gradually disappear, and the width of the transmitted peak gradually narrows, when the $g \gg J$ the system behaves perfect reflection for a single photon.
\\
\\
\subsection{Comparison with the two-level system}

In the following, we compare the single photon scattering properties of the second order nonlinear system with the one-dimensional CRW coupling to a two-level system~\cite{aa}, here we set $U=0$, the current system has the same Hamiltonian form as the two-level system, but the eigenstate of the two system are different, the eigenstate of the latter system can be written as
\begin{eqnarray}
|E_k'\rangle=\sum_ju_k(j)a_j^\dag|0\rangle_a|0\rangle_b|g\rangle + u_b|0\rangle_a|1\rangle_b|g\rangle + u_e|0\rangle_a|0\rangle_b|e\rangle,
\label{11}
\end{eqnarray}
The transmission amplitude is calculated as
\begin{eqnarray}
t'&=&\frac{2i\xi [(E_k - \omega_b)(E_k-\omega_e)-J^2] \sin k}{-g^2(E_k-\omega_e) + 2 i \xi[(E_k-\omega_b)(E_k-\omega_e)-J^2] \sin k},\nonumber\\
\label{12}
\end{eqnarray}

\begin{figure}
\centering
\includegraphics[scale=0.70]{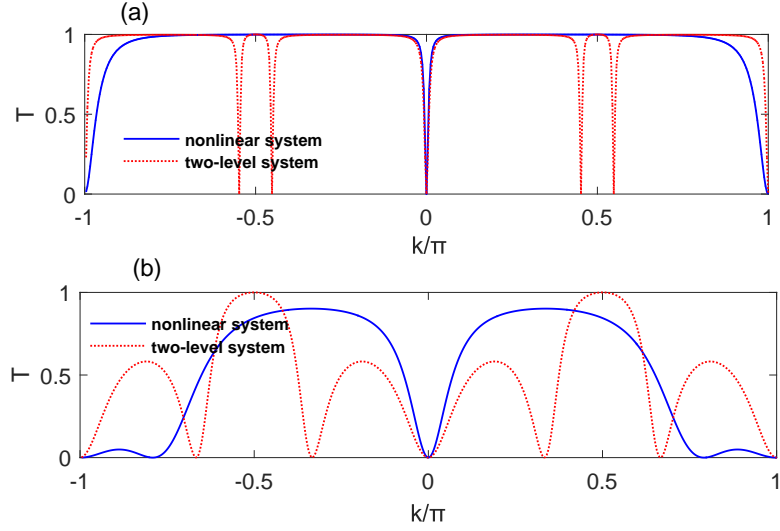}
\caption{(Color online) Transmission rate $T$ as a function of $k$ with different $g$ and $J$. (a) $g=J=0.3$, (b) $g=J=1$. In both (a) and (b), the parameters are $\omega_a=\omega_b=\omega_e=3$ for second-order nonlinear system and the the two-level system. Where the black solid line denotes the second-order nonlinearity system and the red dotted line denotes the two-level system. For convenience all parameters are in units of $\xi$.}
\label{fig5}
\end{figure}

In order to compare the scattering characteristics of two systems under the same coupling conditions, in Fig.~\ref{fig5} we plot the transmission rate $T$ as a function varying with the $k$ for the same $g$ and $J$.

In Fig.~\ref{fig5}(a), we set the coupling strengths $g=J=0.3$, the black solid line denotes the second-order nonlinearity system and the red dotted line denotes the two-level system. We can find that the two system main performance transmission properties, but the second-order nonlinearity system has fewer perfect reflection dips, this is because under the weak coupling strength, the destructive interference caused by second-order nonlinearity is relatively weak, and there is less interference in the transmission path of a single photon, so, the transmission properties are more obvious. In Fig.~\ref{fig5}(b), we set the coupling strength $g=J=1$, and as the coupling strength increases, the reflection control of the system for single photon is enhanced. Different from the reflection dips formed by the two-level system, the second-order nonlinear system forms two perfect reflection regions at $\frac{k}{\pi}=0.8$ to $\frac{k}{\pi}=1$ and $\frac{k}{\pi}=-0.8$ to $\frac{k}{\pi}=-1$, which can better control the single-photon to realize reflection in the perfect reflection areas. So, the current system has the advantage of controlling a single-photon reflection. However, the maximum value of $T$ does not reach $1$ in a large area, which is also a defect of the system.

\section{Scattering properties of a single photon inside a CRW with second-order nonlinearity and Kerr-nonlinearity}

In the current system, because of the resonator $a_0$ and additional cavity $b_0$ are coupled by a second-order nonlinear material, the $g$ is the second-order nonlinear coupling strength, which mediates the conversion of the photon in cavity $a_0$ into two-photon in the cavity $b_0$, so, a Kerr-nonlinear material will have a certain influence on the two-photon in the additional cavity $b_0$, and further affect the scattering properties of a single-photon in the system. In this section we'll discuss how the Kerr nonlinearity affects on this system under the case of $J=0$, which means that the two-level atom is decoupled from the additional cavity $b_0$, the Hamiltonian of this system can be written as
\begin{eqnarray}
H&=&H_a+H_b+H_g+H_U,
\label{13}
\end{eqnarray}
The eigenstate of the current system can be written as
\begin{eqnarray}
|E_k''\rangle=\sum_ju_k(j)a_j^\dag|0\rangle_a|0\rangle_b + u_b|0\rangle_a|2\rangle_b,
\label{11}
\end{eqnarray}
The transmission amplitude is calculated as
\begin{eqnarray}
t''=\frac{2i\xi(E_k-2\omega_b-2U)\sin k}{-2g^2+2i\xi(E_k-2\omega_b-2U)\sin k}.
\label{14}
\end{eqnarray}
where $\Delta_{aU}\equiv E_k - \omega_a$ and $\Delta_{bU}\equiv E_k - 2 \omega_b$, we rewrite the $t''$ of Eq.~(\ref{14}) in another form
\begin{eqnarray}
t''&=&\frac{2i\xi (\Delta_{bU}-2U)\sqrt{1-(\frac{\Delta_{aU}}{2\xi})^2}}{-2 g^2+2i\xi(\Delta_{bU}-2U)\sqrt{1-(\frac{\Delta_{aU}}{2\xi})^2}},
\label{10}
\end{eqnarray}
The corresponding effective potential $V_U$ can be written as
\begin{eqnarray}
V_U=-\frac{-2g^2}{E_k-2\omega_b-2U}.
\label{15}
\end{eqnarray}

In different single-photon control systems, the effective potential is the key to determine the scattering property of the system, and
the scattering process properties of a single incident photon with a specific momentum are mainly determined by the effective potential.

\begin{figure}[h]
\centering
\includegraphics[scale=0.70]{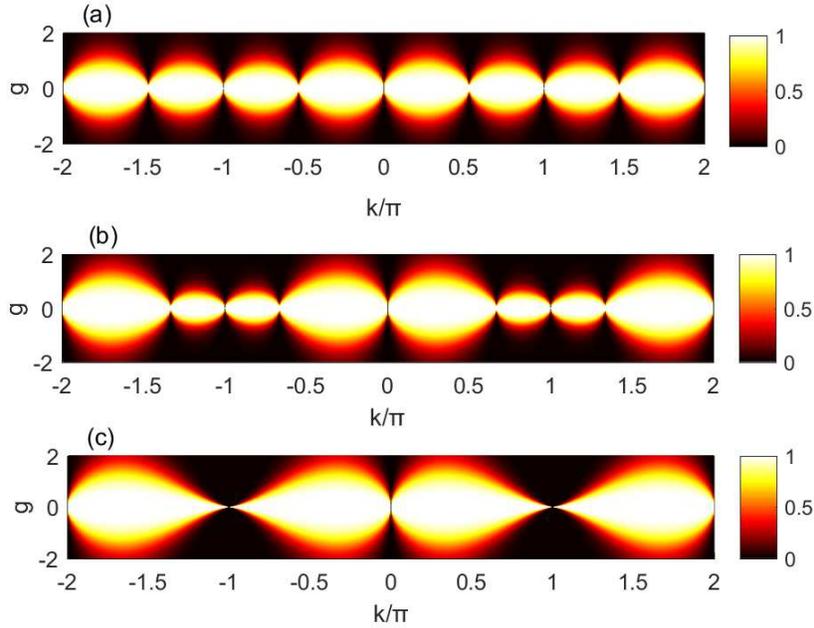}
\caption{(Color online) Transmission rate $T$ as a function of $k$ and $g$ with different $U$. $\omega_a=4$, $\omega_b=2$.(a) $U=0.1$, (b) $U=0.5$ and (c) $U=1$. All the parameters are resealed by the hopping energy $\xi$.}
\label{fig6}
\end{figure}

In Fig.~\ref{fig6}, we plot the transmission rate $T$ as a function of $k$ and $g$ with different $U$, and the parameters $\omega_a=2\omega_b=4$ are same in all the three transmission spectrum.
In Fig.~\ref{fig6}(a), we set the $U=0.1$, and the joint positions of the $T=1$ can be calculated as $k=n\pi$ or $k=\arccos\frac{\omega_a-2(\omega_b+U)}{2\xi}$. If one of the joint position conditions is satisfied, the nodes appear. In Fig.~\ref{fig6}(b), we set $U=0.5$, as the value of $U$ increases, large elliptic transmission areas occurs periodically, and the joint positions of the $T=1$ moves, the role of the Kerr-nonlinearity in the system is similar to that of the two-level atom above. In Fig.~\ref{fig6}(c), we set $U=1$, as the value of $U$ further increase, the joint position conditions $k=\arccos\frac{\omega_a-2(\omega_b+U)}{2\xi}$ doesn't work, the number of nodes reduced, and the transmission area gets bigger obviously. Therefore, the Kerr nonlinearity can also be controlled as a single photon switch in this system.

\begin{figure}
\centering
\includegraphics[scale=0.70]{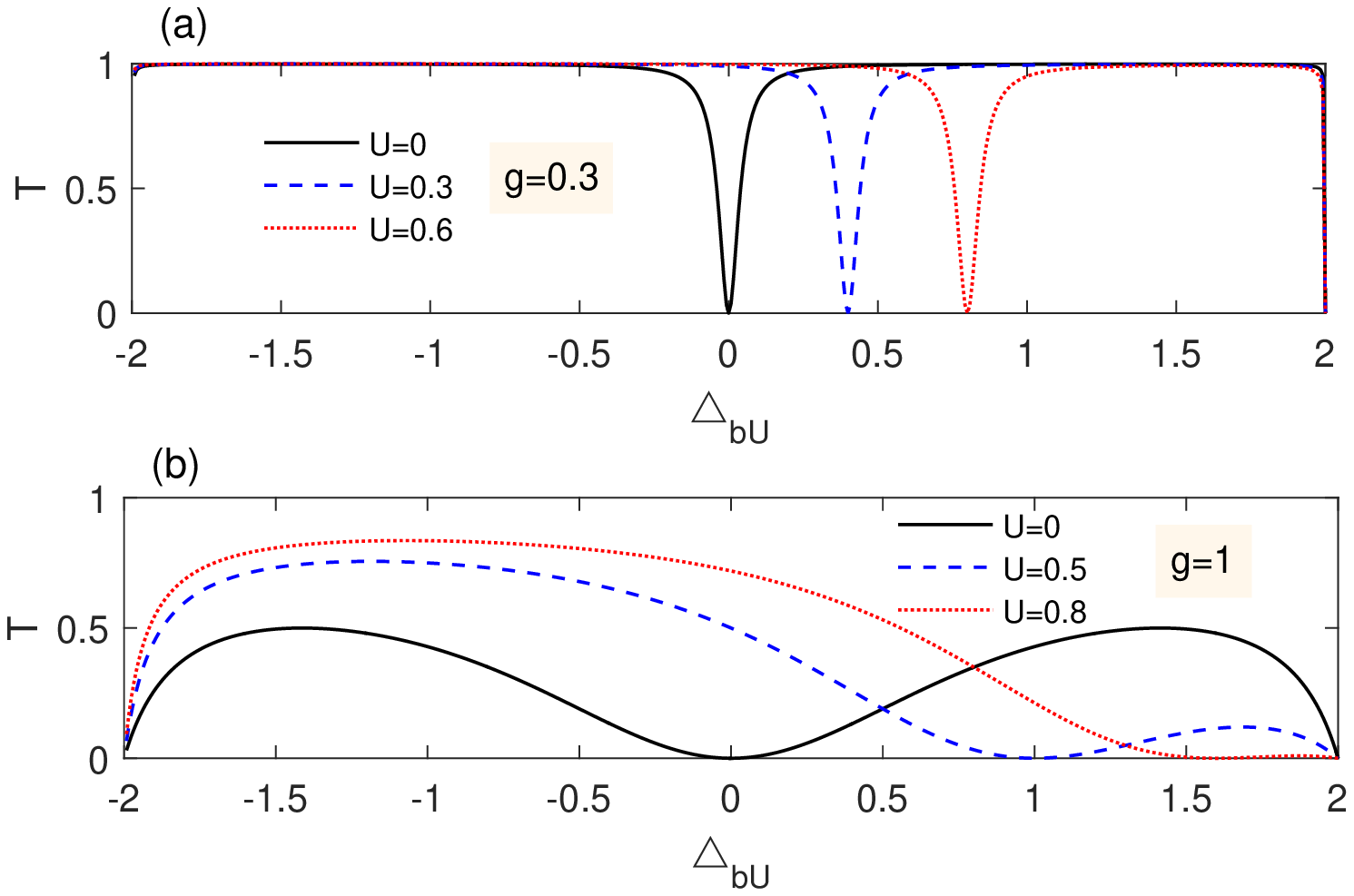}
\caption{(Color online) Transmission rate $T$ as a function of $\Delta_{bU}$ with different $U$. $\omega_a=4$, $\omega_b=2$. (a) $g=0.3$, the black solid line $U=0$, the blue dotted line $U=0.3$ and the red point line $U=0.6$. (b) $g=1$, the black solid line $U=0$, the blue dotted line $U=0.5$ and the red point line $U=0.8$. All the parameters are resealed by the hopping energy $\xi$.}
\label{fig7}
\end{figure}

For further study the effect of Kerr nonlinearity on the scattering properties of the system, in Fig.~\ref{fig6} we plot transmission rate $T$ as a function of $\Delta_{bU}$ with different $U$. In Fig.~\ref{fig7}(a), we set $g=0.3$, and the black solid line $U=0$, the blue dotted line $U=0.3$ and the red point line $U=0.6$, we can find that as the value of $U$ increases the perfect reflection position moves to the right. In Fig.~\ref{fig7}(b), we set $g=1$, and in these three lines we set $U=0$, $U=0.5$ and $U=0.8$, respectively. when $U=0$ a dip appears at $\Delta_{bU}=0$, and the two side peaks are symmetrically distributed, as $U$ increases the dip moves to the right, the maximum value of the left side peak increases and the span widens, and the maximum value of the right side peak decreases and the span narrows, which means that the transmission property of the system is enhanced. When the $U=0.8$, the scattering properties of the system are further altered, and a perfect reflection region is formed in the region from $\Delta_{bU}=1.5$ to $\Delta_{bU}=2$, which increases the system's ability to control the reflection of the single photon.

\subsection{Effects of Kerr-nonlinearity in the hybrid system}
Finally, we discuss the case that in a CRW coupling an additional cavity via second-order nonlinearity, and in the additional cavity both of the two-level atom and the Kerr nonlinearity are excited. The transmission and reflection amplitude equations show in Eqs.~(\ref{09}).

 \begin{figure}[h]
\centering
\includegraphics[scale=0.60]{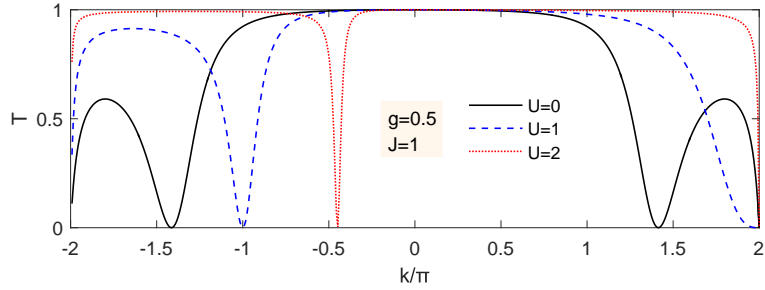}
\caption{(Color online) Transmission rate $T$ as a function of $\frac{k}{\pi}$ with different $U$. The other parameters are $\omega_a=4$, $\omega_b=2$, $g=0.5$ and $J=1$, and the black solid line $U=0$, the blue dotted line $U=1$ and the red point line $U=1$. All the parameters are resealed by the hopping energy $\xi$.}
\label{fig8}
\end{figure}

In Fig.~\ref{fig8}, we set $g=0.5$ and $J=1$, plot the transmission rate $T$ as a function of $\frac{k}{\pi}$ with different Kerr-nonlinear strength $U$. In the black solid line $U=0$, there were two symmetrical dips in the transmission spectrum, and the duty cycle of the two side peaks are the same. When the $U=1$, the two dips moves to the right, the maximum value and  span of the left side peak increases obviously, the transmission properties increases of the single-photon in the system, the results show in the blue dotted line. As the $U$ further increase, the dip areas has narrowed significantly, the whole system presents obvious transmission property. Based on the above conclusions, when second-order nonlinear strength $g$ and the coupling strength of two-level atom $J$ are determined, The Kerr nonlinear strength $U$ can be used as a single photon switch to adjust the scattering property of a single photon in the system.

\section{Conclusion}
We have investigated the coherent controlling of a single photon transport in a
CRW coupled to an additional resonator via the second-order nonlinearity, where the additional resonator is embedded with two-level atom and filled with Kerr-nonlinear materials. By the discrete coordinates approach we obtain the reflection and transmission rate for the single-photon propagating in this system. The effects of second-order nonlinearity, Kerr nonlinearity and two-level atoms acting as a switch to control the transmission and reflection of single photons in a one-dimensional coupled resonant waveguide are analyzed, respectively. The results shown that the single photon can be perfectly reflected or transmitted by tuning the the system parameters.
By compared with the two-level system we find that under the same parameters the current system it easier to control a single photon to realize perfect reflection, so as to better realize the function of single-photon switch. The Kerr nonlinearity can also be used as a single photon switch to control the single-photon scattering properties in the system. The hybrid system will provide more ways for the realization of quantum devices and promote its development.

\section*{Funding}
This work is supported by the National Natural Science Foundation of China with Grants No. 11647054, the Science and Technology Development Program of Jilin province, China with Grant No. 2018-0520165JH, the Jiangxi Education Department Fund under Grant No. GJJ180873.

\section*{Disclosures:} The authors declare no conflicts of interest.\\


\begin{thebibliography}{9}
\bibitem{aa} M. T. Cheng, X. S. Ma, M. T. Ding, Y. Q. Luo, and G. X. Zhao, ``Single-photon transport in one-dimensional coupled-resonator waveguide with local and nonlocal coupling to a nanocavity containing a two-level system", Phys. Rev. A \textbf{85}, 053840 (2012).
\bibitem{1}O. Astafiev, A. M. Zagoskin, A. A. Abdumalikov Jr. Y. A. Paskin, T. Yamamoto, K. Inomata, Y. Nakamura, and J. S. Tsai, ``Resonance fluorescence of a single artificial atom", Science \textbf{327}, 840-843 (2010).
\bibitem{2}B. Lounis and M. Orrit, ``Single-photon sources", Rep. Prog. Phys. \textbf{68}, 1129šC1179 (2005).
\bibitem{3}M. Hijlkema, B. Weber, H. P. Specht, S. C. Webster, A. Kuhn, and G. Rempe, ``A single-photon server with just one atom", Nature Physics. \textbf{3}, 253šC255 (2007).
\bibitem{4}S. Buckley, K. Rivoire, and J. V\u{u}ckov'c, ``Engineered quantum dot single-photon sources", Rep. Prog. Phys. \textbf{75}, 126503 (2012).
\bibitem{5}J.-T. Shen and S. Fan, ``Coherent Single Photon Transport in A One-Dimensional Waveguide Coupled with Superconducting Quantum Bits", Phys. Rev. Lett. \textbf{95}, 213001 (2005).
\bibitem{6}D. E. Chang, A. S. Sorensen, E. A. Demler, and M. D. Lukin, ``A single-photon transistor using nanoscale surface plasmons", Nat. Phys. \textbf{3}, 807 (2007).
\bibitem{7}J.-T. Shen and S. Fan, ``Strongly Correlated Two-Photon Transport in A One-Dimensional Waveguide Coupled to A Two-Level System", Phys. Rev. Lett. \textbf{98}, 153003 (2007).
\bibitem{8}L. Zhou, Z. R. Gong, Y.-x. Liu, C. P. Sun, and F. Nori, ``Controllable Scattering of a Single Photon Inside a One-Dimensional Resonator Waveguide", Phys. Rev. Lett. \textbf{101}, 100501 (2008).
\bibitem{9}D. E. Chang, L. Jiang, A. V. Gorshkov, and H. J. Kimble,``Cavity QED with atomic mirrors", New J. Phys. \textbf{14}, 063003 (2012).
\bibitem{10} Y. Chang, Z. R. Gong, and C. P. Sun, ``Multiatomic mirror for perfect refiection of single photons in a wideband of frequency", Phys. Rev. A \textbf{83}, 013825 (2011).
\bibitem{11}L. Zhou, H. Dong, Y.-x. Liu, C. P. Sun, and F. Nori, ``Quantum supercavity with atomic mirrors", Phys. Rev. A \textbf{78}, 063827 (2008).
\bibitem{12}T. G. Tiecke, J. D. Thompson, N. P. de Leon, L. R. Liu, V. Vuletic, and M. D. Lukin, ``Nanophotonic quantum phase switch with a single atom", Nature (London) \textbf{508}, 241 (2014).
\bibitem{13}H. Zoubi and H. Ritsch, ``Hybrid quantum system of a nanofiber mode coupled to two chains of optically trapped atoms", New J. Phys. \textbf{12}, 103014 (2010).
 \bibitem{14}S. Okaba, T. Takano, F. Benabid, T. Bradley, L. Vincetti, Z. Maizelis, V. Yampol¡¯skii, F. Nori, and H. Katori, ``Lamb-Dicke spectroscopy of atoms in a hollow-core photonic crystal fibre", Nat. Commun. \textbf{5}, 4096 (2014).
\bibitem{15}J. T. Shen and S. Fan, ``Coherent Single Photon Transport in a One-Dimensional Waveguide Coupled with Superconducting Quantum Bits",  Phys. Rev. Lett. \textbf{95}, 213001 (2005).
\bibitem{16}L. Zhou, Z. R. Gong, Y. X. Liu, C. P. Sun, and F. Nori, ``Controlling the transport of single photons by tuning the frequency of either one or two cavities in an array of coupled cavities", Phys. Rev. Lett. \textbf{101}, 100501 (2008).
\bibitem{17}J. T. Shen and S. Fan, ``Theory of single-photon transport in a single-mode waveguide. I. Coupling to a cavity containing a two-level atom", Phys. Rev. A \textbf{79}, 023837 (2009).
\bibitem{18}W. Z. Jia and Z. D. Wang, ``Single-photon transport in a one-dimensional waveguide coupling to a hybrid atom-optomechanical system", Phys. Rev. A \textbf{88}, 063821 (2013).
\bibitem{19}L. Neumeier, M. Leib, and M. J. Hartmann, ``Single-Photon Transistor in Circuit Quantum Electrodynamics",  Phys. Rev. Lett. \textbf{111}, 063601 (2013).
\bibitem{20}W. Qin and F. Nori, ``Controllable single-photon transport between remote coupled-cavity arrays",  Phys. Rev. A \textbf{93}, 032337 (2016).
\bibitem{21}I.-C. Hoi, C. M. Wilson, G. Johansson, T. Palomaki, B. Peropadre,and P.Delsing, ``Demonstration of a Single-Photon Router in the Microwave Regime", Phys. Rev. Lett. \textbf{107},073601(2011).
\bibitem{22}B. Dayan, A. S. Parkins, T. Aoki, E. P. Ostby, K. J. Vahala, and H. J. Kimble, ``A Photon Turnstile Dynamically Regulated by One Atom",  Science \textbf{319}, 1062 (2008).
\bibitem{23}A. V. Akimov, A. Mukherjee, C. L. Yu, D. E. Chang, A. S. Zibrov, P. R. Hemmer, H. Park, and M. D. Lukin, ``Generation of single optical plasmons in metallic nanowires coupled to quantum dots", Nature (London) \textbf{450}, 402 (2007).
\bibitem{24}A. Goban, C. L. Hung, S. P. Yu, J. D. Hood, J. A. Muniz, J. H. Lee, M. J. Martin, A. C. McClung, K. S. Choi, D. E. Chang, O. Painter, and H. J. Kimble, ``Atom-light interactions in photonic crystals", Nat. Commun. \textbf{5}, 3808 (2014).
\bibitem{25}C.-L.Hung,S.M.Meenehan,D.E.Chang,O.Painter,andH.J. Kimble, ``Trapped atoms in one-dimensional photonic crystals", New J. Phys. \textbf{15}, 083026 (2013).
\bibitem{26}A. Goban, C.-L. Hung, J. D. Hood, S.-P. Yu, J. A. Muniz, O. Painter, and H. J. Kimble, ``Superradiance for Atoms Trapped Along a Photonic Crystal Waveguide", Phys. Rev. Lett. \textbf{115}, 063601 (2015).
 \bibitem{27}R. Mitsch, C. Sayrin, B. Albrecht, P. Schneeweiss, and A. Rauschenbeutel, ``Quantum state-controlled directional spontaneous emission of photons into a nanophotonic waveguide", Nat. Commun. \textbf{5}, 5713 (2014).
\bibitem{28}E. Vetsch, D. Reitz, G. SagušŠ, R. Schmidt, S. T. Dawkins, and A. Rauschenbeutel, ``Optical Interface Created by Laser-Cooled Atoms Trapped in the Evanescent Field Surrounding an Optical Nanofiber", Phys. Rev. Lett. \textbf{104}, 203603 (2010).
\bibitem{29}A. Goban, K. S. Choi, D. J. Alton, D. Ding, C. Lacro?te, M. Pototschnig, T. Thiele, N. P. Stern, and H. J. Kimble, ``Demonstration of A State-Insensitive, Compensated Nanofiber Trap", Phys. Rev. Lett. \textbf{109}, 033603 (2012).
\bibitem{30}R. Yalla, M. Sadgrove, K. P. Nayak, and K. Hakuta, ``Cavity Quantum Electrodynamics On a Nanofiber Using a Composite Photonic Crystal Cavity", Phys. Rev. Lett. \textbf{113}, 143601 (2014).
\bibitem{31}G. Epple, K. S. Kleinbach, T. G. Euser, N. Y. Joly, T. Pfau, P. St. J. Russell, and R. L\"{o}w, ``Rydberg atoms in hollow-core photonic crystal fibres", Nat. Commun. \textbf{5}, 4132 (2014).
 \bibitem{32}M. Langbecker, M. Noaman, N. Kj\"{o}rgaard, F. Benabid, and P. Windpassinger, ``Rydberg excitation of cold atoms inside a hollow core fiber", arXiv: 1706.07666.
 \bibitem{33}P. Solano, P. Barberis-Blostein, F. K. Fatemi, L. A. Orozco, and S. L. Rolston, ``Super-radiance reveals in?nite-range dipole interactions through a nanofiber", arXiv: 1704.07486.
 \bibitem{34}J.Javanainen,J.Ruostekoski,B.Vestergaard,andM.R.Francis, ``One-dimensional modelling of light propagation in dense and degenerate samples", Phys. Rev. A \textbf{59}, 649 (1999).
\bibitem{35}F. Le Kien and K. Hakuta, ``Slowing down of a guided light ?eld along a nanofiber in a coldatomicgas", Phys. Rev. A \textbf{79},013818 (2009).
\bibitem{36}F. L. Kien and K. Hakuta, ``Cooperative enhancement of channeling of emission from atoms into a nanofiber", Phys. Rev. A  \textbf{77}, 013801 (2008).
\bibitem{37}Z. Liao, X. Zeng, S.-Y. Zhu, and M. S. Zubairy, ``Single-photon transportthroughanatomicchaincoupledtoaone-dimensional nanophotonic waveguide", Phys. Rev. A \textbf{92}, 023806 (2015).
\bibitem{38}J. Ruostekoski and J. Javanainen, Emergence of ``Correlated Optics in One-Dimensional Waveguides for Classical and Quantum Atomic Gases", Phys. Rev. Lett. \textbf{117}, 143602 (2016).
\bibitem{39}M. T. Manzoni, D. E. Chang, and J. S. Douglas, ``Simulating quantum light propagation through atomic ensembles using matrix product states", arXiv: 1702.05954.
\bibitem{40}A. Asenjo-Garcia, J. D. Hood, D. E. Chang, and H. J. Kimble, ``Atom-light interactions in quasi-one-dimensional nanostructures: A Green¡¯s-function perspective", Phys. Rev. A \textbf{95}, 033818 (2017).
\bibitem{45}A. Faraon, E. Waks, D. Englund, I. Fushman, and J. Vu¡Šckovi¡ä c, ``Efficient photonic crystal cavity-waveguide couplers", Appl. Phys. Lett. \textbf{90}, 073102 (2007).
 \bibitem{46}A. Wallraff, D. I. Schuster, A. Blais, L. Frunzio, R.-S. Huang, J. Majer, S. Kumar, S. M. Girvin, and R. J. Schoelkopf, ``Strong coupling of a single photon to a superconducting qubit using circuit quantum electrodynamics",  Nature (London) \textbf{431}, 162 (2004).
 \bibitem{47}I. C. Hoi, C. M. Wilson, G. Johansson, T. Palomaki, B. Peropadre, and P. Delsing, ``Demonstration of a Single-Photon Router in the Microwave Regime", Phys. Rev. Lett. \textbf{107}, 073601 (2011).
 \bibitem{48}L. Zhou, Y. B. Gao, Z. Song, and C. P. Sun, ``Coherent output of photons from coupled superconducting transmission line resonators controlled by charge qubits", Phys. Rev. A \textbf{77}, 013831 (2008).
\bibitem{49}A. E. Miroshnichenko, S. F. Mingaleev, S. Flach, and Y. S. Kivshar, ``Nonlinear Fano resonance and bistable wave transmission", Phys. Rev. E \textbf{71}, 036626 (2005).
\bibitem{50}A. E. Miroshnichenko, ``Nonlinear Fano-Feshbach resonances", Phys. Rev. E \textbf{79}, 026611 (2009).
\bibitem{51}A. E. Miroshnichenko, S. Flach, and Y. S. Kivshar, ``Fano resonances in nanoscale structures", Rev. Mod. Phys. \textbf{82}, 2257 (2010).
\bibitem{l1}S. Das, V. E. Elfving, F. Reiter, and A. S. S{\o}rensen,``Photon scattering from a system of multilevel quantum emitters. I. Formalism", Phys. Rev. A \textbf{97}, 043837 (2018).
\bibitem{l2}S. Das, V. E. Elfving, F. Reiter, and A. S. S{\o}rensen, ``Photon scattering from a system of multilevel quantum emitters. II.
Application to emitters coupled to a one-dimensional waveguide", Phys. Rev. A \textbf{97}, 043838 (2018).
\bibitem{52}S. Das, V. E. Elfving, F. Reiter, and A. S. S?rensen,``Photon scattering from a system of multilevel quantum emitters. I. Formalism", Phys. Rev. A \textbf{97}, 043837 (2018)
\bibitem{53}A. Rodriguez, M. Soljacic, J. D. Joannopoulos, and S. G. Johnson, ``$\chi^{(2)}$ and $\chi^{(3)}$ harmonic generation at a critical power in inhomogeneous doubly resonant cavities", Opt. Express \textbf{15}, 7303-7318 (2007)
\bibitem{54}Z.-F. Bi, A. W. Rodriguez, H. Hashemi, D. Duchesne, M. Loncar, K.-M. Wang, and S. G. Johnson, ``High-
efficiency second-harmonic generation in doubly-resonant $\chi^{(2)}$ microring resonators", Opt. Express \textbf{20}, 7526-7543
(2012).
\bibitem{55}55. P. S. Kuo, J. Bravo-Abad, and G. S. Solomon, ``Second-harmonic generation using-quasi-phasematching in a
GaAs whispering-gallery-mode microcavity", Nat. Commun. \textbf{5}, 3109 (2014).
\bibitem{56}W. T. M. Irvine, K. Hennessy, and D. Bouwmeester, ``Strong Coupling between Single Photons in Semiconductor
Microcavitie", Phys. Rev. Lett. \textbf{96}, 057405 (2006).
\bibitem{ll1}C. Wang, X. Xiong, N. Andrade, V. Venkataraman, X.-F. Ren, G.-C. Guo, and Ma. Lon\v{c}ar, "Second harmonic generation in nano-structured thin-film lithium niobate waveguides," Opt. Express \textbf{25}, 6963-6973 (2017).
\bibitem{ll2}N. Sinclair, D. Oblak, C. W. Thiel, R. L. Cone, and W. Tittel, ``Properties of a Rare-Earth-Ion-Doped Waveguide at Sub-Kelvin Temperatures for Quantum Signal Processing",
Phys. Rev. Lett. \textbf{118}, 100504 (2017).
\bibitem{ll3}K. Wang, J. Li, R. Gao, and Z. Qi, "LiNbO3 waveguide Based Fourier Transform Spectrometer with Algorithmic Enhancement of Spectral Resolution," in Advanced Photonics 2018 (BGPP, IPR, NP, NOMA, Sensors, Networks, SPPCom, SOF), OSA Techncail Digest (online) (Optical Society of America, 2018), paper JTu6D.1.
\bibitem{57}A. Majumdar and D. Gerace, ``Single-photon blockade in doubly resonant nanocavities with second-order nonlinearity", Phys. Rev. B \textbf{87}, 235319 (2013).
\bibitem{58}Y. H. Zhou, H. Z. Shen, and X. X. Yi, ``Unconventional photon blockade with second-order nonlinearity", Phys. Rev. A \textbf{92}, 023838 (2015).
\bibitem{59}Y. H. Zhou, H. Z. Shen, X. Y. Luo, Y. Wang, F. Gao, and C. Y. Xin, ``Tunable three-wave-mixing-induced transparency", Phys. Rev. A \textbf{96}, 063815 (2017).
\bibitem{60}Y. H. Zhou, S. S. Zhang, H. Z. Shen, and X. X. Yi, ``Second-order nonlinearity induced transparency", Opt. Lett. \textbf{42}, 1289-1292 (2017).
\bibitem{ff1}Z. H. Wang, C. P. Sun, and Yong Li, ``Microwave degenerate parametric down-conversion with a single cyclic
three-level system in a circuit-QED setup", Phys. Rev. A \textbf{91}, 043801 (2015).
\bibitem{shiyan0} M. Sandberg, C. M. Wilson, F. Persson, T. Bauch, G. Johansson, V. Shumeiko, T. Duty, and P. Delsing, Tuning the field in a microwave resonator faster than the photon lifetime, Appl. Phys. Lett. \textbf{92}, 203501 (2008).


%
%
%
%
%
%
%
%
%
%
%

\end{thebibliography}
\end{document}